\documentclass[twocolumn,aps,prl]{revtex4-1}

\usepackage{epsfig}
\usepackage[usenames]{color}
\usepackage{bm}
\usepackage{amssymb}

\newcommand{\nn}{\nonumber}

\newcommand{\ve} {\varepsilon}

\newcommand{\be}{\begin{eqnarray}}
\newcommand{\ee}{\end{eqnarray}}
\newcommand{\la}{\langle}
\newcommand{\ra}{\rangle}
\newcommand{\rar}{\rightarrow}
\newcommand{\da}{\downarrow}
\newcommand{\ua}{\uparrow}

\begin{document}

\title{Quantum annealing and thermalization: insights from integrability}

\author {Fuxiang Li}
\affiliation{School of Physics and Electronics, Hunan University, Changsha 410082, China}
\affiliation{Theoretical Division, Los Alamos National Laboratory, B213, Los Alamos, NM 87545}
\affiliation{Center for Nonlinear Studies, Los Alamos National Laboratory, Los Alamos, NM 87545}
\author{Vladimir Y. Chernyak}
\affiliation{Department of Chemistry and Department of Mathematics, Wayne State University, 5101 Cass Ave, Detroit, Michigan 48202, USA}

\author{Nikolai A. Sinitsyn}
\affiliation{Theoretical Division, Los Alamos National Laboratory, B213, Los Alamos, NM
87545}
\date{\today}

\begin{abstract}
We solve a model that has basic features that {are desired for quantum annealing computations: entanglement in the ground state,}  controllable annealing speed, ground state energy separated by a gap during the whole evolution, and programmable computational problem that is encoded by parameters of the Ising part of the spin Hamiltonian. Our solution enables exact nonperturbative characterization of  final nonadiabatic excitations, including scaling of their number with the annealing rate and the system size. {We prove that quantum correlations can accelerate computations and, at the end of the annealing protocol, lead to the perfect Gibbs distribution of all microstates. }
 \end{abstract}


\date{\today}

\maketitle

\begin{figure} [t]
{\includegraphics[width=\columnwidth]{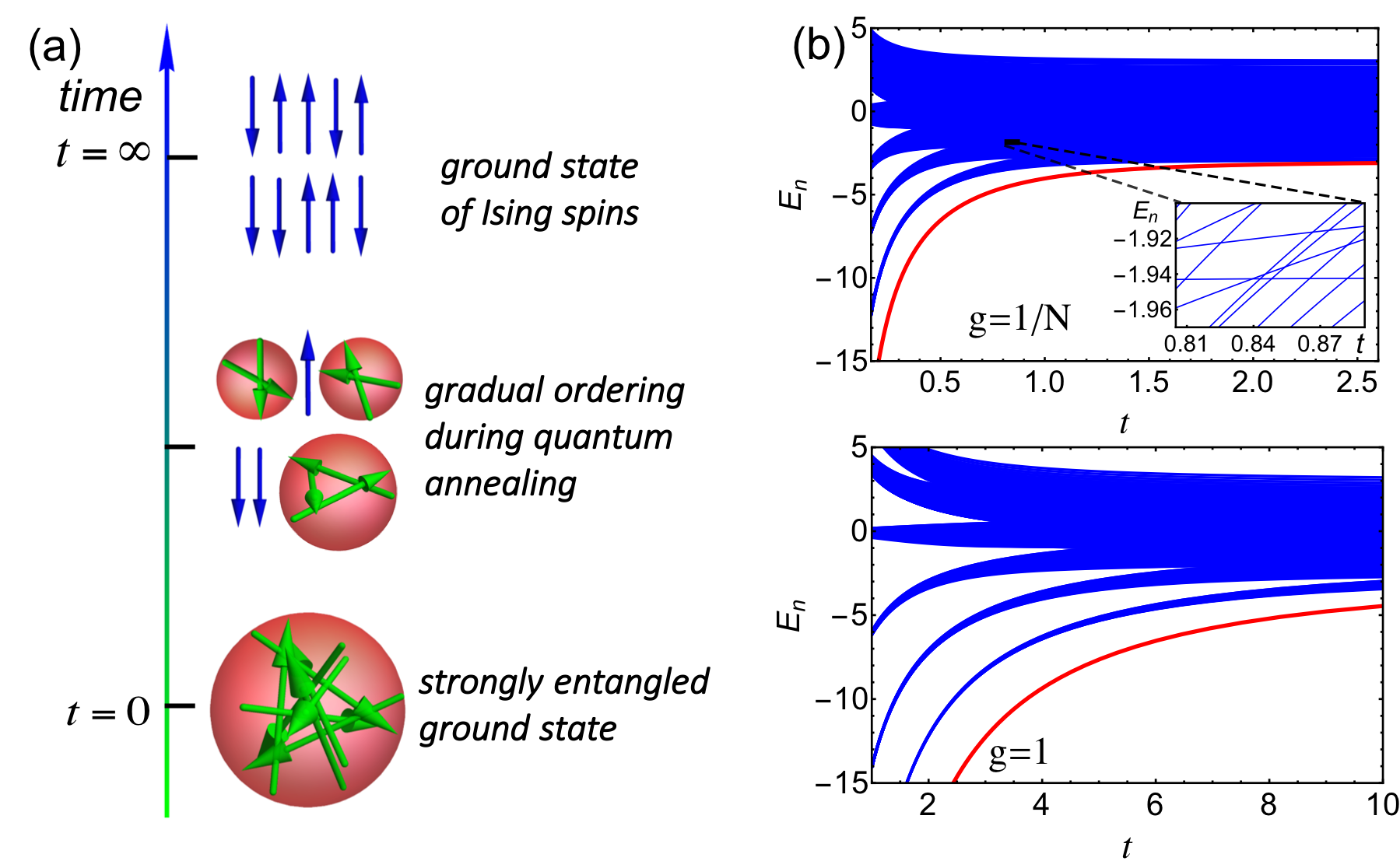}}
\caption{{\bf (a)} {During QA, the  entangled ground state is  transformed adiabatically into the  ground state of the Ising spin Hamiltonian. }
{\bf (b)} Evolution of the spectra of the QA Hamiltonian (\ref{h-bcs0}) in nonadiabatic ($g=1/N$, top) and nearly-adiabatic ($g=1$, bottom) regimes.
 The  ground level is marked by red color.
  Here, $N=12$, $S_{\rm tot}^z=0$, $\ve_j=j/N + \xi_j$, and $\xi_j$ are uniformly distributed random real numbers from the range $(-1/(2N), 1/(2N))$. The inset shows exact level crossings indicating model's integrability.
   }
\label{spectra}
\end{figure}

Many optimization problems can be reformulated in terms of searching for a  configuration that minimizes a Hamiltonian  ${H}_A({s}_1,\ldots, {s}_{N})$ of $N$ Ising spins $s_j$ \cite{finnila1994quantum,santoro2002theory, das2008colloquium}. This task is often so complex that it cannot be solved with modern computers.
 The idea of quantum annealing (QA) is to treat the Ising spins as $z$-components of quantum spins-1/2, $ \hat{\bm s}_j$, and realize quantum evolution with a Hamiltonian
\be
\hat{H}(t) = \hat{H}_A(\hat{s}^z_1,\ldots, \hat{s}^z_{N}) +g(t)\hat{H}_B(\hat{\bm s}_1,\ldots, \hat{\bm s}_{N}),
\label{aham1}
\ee
where $\hat{H}_B$ has a ground state that  { overlaps }with all possible QA outcomes and  does not discriminate against some of them at the start.
Parameter $g(t)$ is  large at $t= 0$ but decays to zero at $t\rar \infty$. According to the adiabatic theorem, a system that is initially in the ground state  remains in the instantaneous ground state if  the lowest energy is always nondegenerate  and parameters change sufficiently slowly.
So,  as we illustrate in Fig.~\ref{spectra}(a), slow decay of $g(t)$ converts the ground state of $\hat{H}_B$   into the ground state of $\hat{H}_A$,
 which is then read  by measuring  spins along the $z$-axis.

  In practice,  the annealing time is restricted, so nonadiabatic excitations become inevitable \cite{annealing-time1,annealing-time2,annealing-time3, barends2016digitized}.
Nevertheless,  at $N\gg1$, there are optimization problems with some error  tolerance.
In this letter, we solve a minimal model of QA and show that:

 (i) tolerance of a computational goal to a small number of  errors allows QA protocols {that introduce extra quantum correlations in order to } reduce the required
computation time  by a factor  $\sim 1/N$ in comparison to the conventionally justified QA time.

(ii) the distribution of nonadiabatic excitations in a closed quantum system after QA can be completely thermalized;

 (iii) this  thermalization is encoded in  integrability, i.e., the possibility to describe  the behavior analytically.

 The first property justifies the error-tolerant QA computation technology, the second one proves that  averaging over unknown conditions is not needed to find thermalization in coherent evolution, and the third one counters the common belief, taking roots in the numerical experiment by Fermi-Pasta-Ulam-Tsingou \cite{fermi1955studies},  that complete thermalization is incompatible with integrability.

{Initial quantum correlations are not required for QA but our goal here is to learn if they can be a resource for accelerated computations.}
The simplest Hamiltonian of $N$ spins with entanglement in the ground state is all-to-all coupling \cite{lechner2015quantum, mcmahon2016fully,nigg2017robust}, $\hat{H}_B = -\sum_{i\ne j}^{N} \hat{s}_i^+ \hat{s}_j^-$,  restricted to a sector with a conserved total spin. The ground state of $\hat{H}_B$ is the  sum of all eigenstates of $\hat{H}_A$ with the same  ${S}^{z}_{\rm tot}=\sum_{j=1}^{N} {s}^z_j $:
\be
|\psi_0 \ra \sim |\ua \ua \ldots \da \da \ra+ |\ua \da \ldots \ua \da \ra+\ldots +|\da \da \ldots \ua \ua \ra.
\label{gs1}
\ee
The simplest to write  QA protocol is the inverse time decay, $g(t)=g/t$, where $t\in (0_+,\infty)$, $g$ is a  constant; and
the simplest Ising Hamiltonian is  $\hat{H}_A = \sum_{j=1}^{N} \varepsilon_j \hat{s}^z_j$, where the vector of constant parameters, ${\bm \ve}=(\ve_1,\ldots, \ve_{N})$, is  programmable for computations. So, the minimal QA Hamiltonian for our goals is
\be
 \hat{H}_{\rm BCS}(t) \!=\! \sum_{j} \ve_j \hat{s}^z_j \!-\!\frac{g}{t} \sum_{j\ne k} \hat{s}^+_j \hat{s}^-_k, \,\,\,\,\, j,k=1,\ldots, N.
\label{h-bcs0}
\ee
Let ${S}^{z}_{\rm tot}=0$ and all constants $\varepsilon_j$ be nondegenerate. The ground state of $H_A$ has   then $N/2$ spins   down and $N/2$ spins  up; all down-spins have larger $\varepsilon_j$  than  all up-spins.
Hence, QA with $\hat{H}_{\rm BCS}$ solves an array sorting problem: to find $N/2$ indices $j$ that mark the largest $\varepsilon_j$.

The time-{\it in}dependent version of $\hat{H}_{\rm BCS}$ is  equivalent to the Bardeen-Cooper-Schrieffer model  of superconductivity \cite{andersonbcs}.  Its nonequilibrium dynamics has attracted considerable interest both experimentally \cite{matsunaga2012nonequilibrium,matsunaga2013higgs} and theoretically \cite{yuzbashyan2015quantum,bcs-theory}. Recently, the time-dependent model (\ref{h-bcs0}) was proved to be integrable \cite{sinitsyn2017integrable}, { but its solution for arbitrary $t$ in terms of repeated contour integrals \cite{yuzbahsyan-10annp} is too complex  to reveal physical properties of QA.  Therefore, here we will develop
a different approach that will target the desired characteristics directly.}

Deviation from adiabaticity is controlled continuously
 in $ \hat{H}_{\rm BCS}(t)$, as shown   in Fig.~\ref{spectra}(b): the ground level is always separated by a gap from the rest of the spectrum but approaches other levels slower when $g$ is larger.
Precision of  QA is usually characterized by the probability $P_G$ to remain in the ground state at $t\rar\infty$. According to the Landau-Zener formula, $P_G$ is determined by the size of the  energy distance $\Delta$ to the nearest energy level and the characteristic rate   $\beta$ with which this gap changes:
$
 P_{LZ}=1-e^{-2\pi \Delta^2/\beta}.
$
 At $t\rar \infty$, the ground level  of $\hat{H}_{\rm BCS}$ is separated from the lowest energy excitation by  $\Delta = |\varepsilon_i-\epsilon_j|$, where $i$ and $j$ are indexes of spins for which this energy difference is minimal. Coupling between these spins becomes comparable to $\Delta$ at the effective annealing time $\tau \sim g/\Delta$, and the characteristic rate with which this coupling changes is $\beta =|d(g/t)/dt|_{t=\tau} = \Delta^2/g$. This leads to the rough estimate in the adiabatic limit: $ P_G\sim  1-e^{-2\pi g}$, which we confirm in Fig.~2{(a) by comparing to numerical results. Hence,  values $g>1$ correspond to  adiabatic QA.

To understand the regime at $g<1$, we assume in what follows that  $0<\ve_1<\ve_2<\ldots<\ve_{N}$, and introduce a new accuracy characteristic:
\be
\eta\equiv (4/N) \sum_{k=1}^{N/2}  {s}^z_k,
\ee
where $s^z_k$ is the outcome of the $k$-th spin polarization measurement. The ground state of $\hat{H}_A$ at $S_{\rm tot}^z=0$ has $\eta=1$.  Excitations reduce $\eta$, e.g., $\eta=0$ means complete loss of valuable information.

\begin{figure} [t]
{\includegraphics[width=0.96\columnwidth]{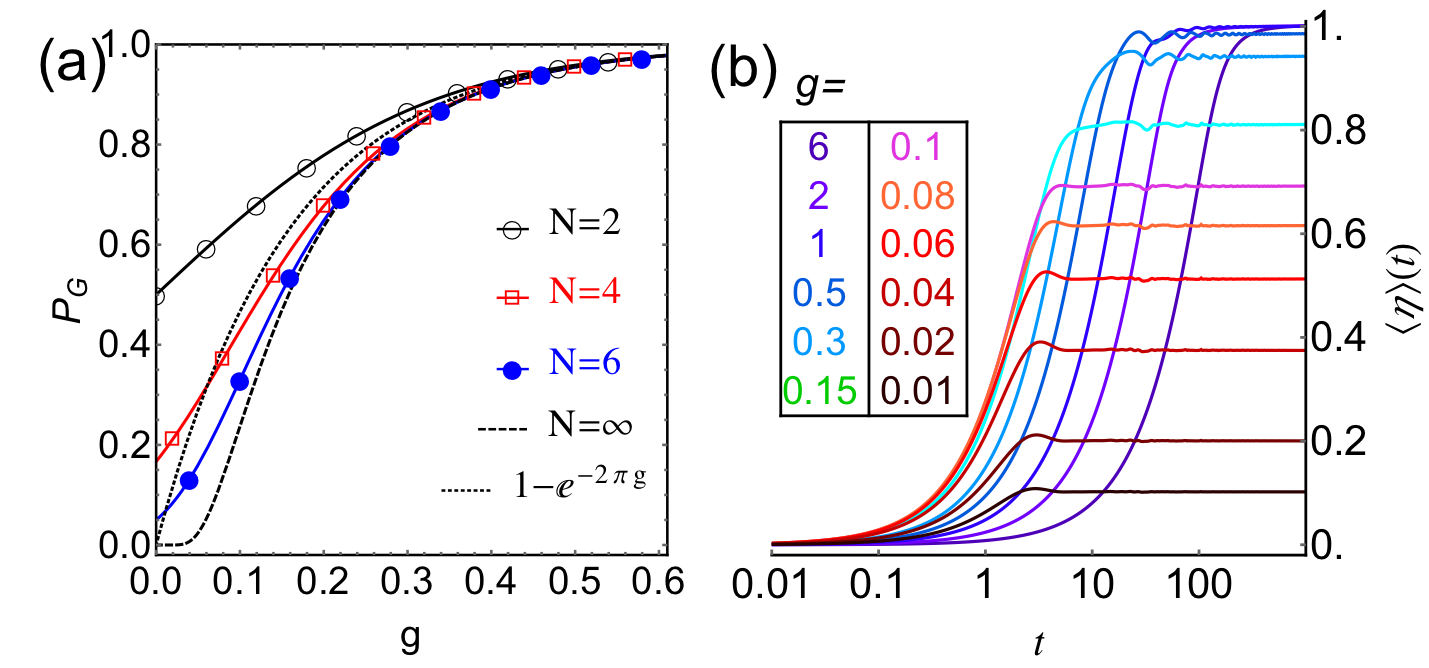}}
\caption{ {\bf (a)}  The probabilities to remain in the ground state at different $g$ and $N$. Solid curves  and  the limit $N\rar \infty$ (black dashed curve)
 are predictions of Eq.~(\ref{eq:Psjf}) and point markers are the numerical results  \cite{supp}.
{\bf (b)} Time-dependence of computation accuracy. Solid curves are results of the numerical solution for the Hamiltonian $\hat{H}_{\rm BCS}(t)$ with $N=12$, $S^z_{\rm tot}=0$, and the same $\ve_j$ as in Fig.~\ref{spectra}(b).}
 \label{fig:etat}
\end{figure}
In Fig.~\ref{fig:etat}(b) we show time-dependence of the mean value $\la \eta \ra$ at different $g$, obtained  by solving  the Schr\"odinger equation with $\hat{H}_{\rm BCS}$ for $N=12$ numerically. Saturation of $\la \eta \ra$ means that one can interrupt evolution at finite $t$ without loosing accuracy. Final $\la \eta \ra$  is growing with $g$ and at  $g=1/N$ it reaches values $\la \eta \ra > 0.6$, at which over  80\% of spins  point correctly along their ground state directions.  At $g<1/N$, the time to saturation is mostly defined by the energy parameters $\ve_j$ and almost does not change with $g$. For $g>1/N$,  this time is growing and becomes   about a factor $N$  longer at $g=1$ than at $g=1/N$, in agreement with our rough  estimate $\tau\sim g/\Delta$.
Figure~\ref{fig:etat}(b)  also suggests that $\la \eta \ra = 1-O(1/N)$ is reached at values of $g$ outside the adiabatic regime.  However, numerical simulations are not decisive here because they are restricted to small $N$. So, we will develop an  analytical approach that will confirm this expectation.

To understand behavior at arbitrary $N$, we recall that $\hat{H}_{\rm BCS}$ commutes with  $N$ Gaudin Hamiltonians \cite{yuzbashyan2005integrable}:
\be
\nonumber \hat{H}_j=t \hat{s}_j^z - 2g \sum \limits_{k\ne j} \frac{\hat{\bf s}_j \cdot \hat{\bf s}_k}{\ve_j-\ve_k},\quad k,j=1,\ldots, N,
\label{c-bcs1}
\ee
which also satisfy conditions:
$\partial_{\ve_j} \hat{H}_{\rm BCS} = \partial_t \hat{H}_j$ and $\partial_{\ve_j} \hat{H}_{i} = \partial_{\ve_i} \hat{H}_j$ for all $i,j$.
{According to \cite{sinitsyn2017integrable}, this property is what makes the model (\ref{h-bcs0}) analytically solvable.} Following \cite{sinitsyn2017integrable},
we introduce multi-time vector ${\bm t}$,  where $t^0\equiv t$, $t^j \equiv \varepsilon_j$ and write an operator of evolution in this multi-time space
\be
\nonumber \hat{U}=\hat{\cal T} {\rm exp}\left[-i\int_{\cal{P}} \sum_{\mu=0}^{N} \hat{H}_{\mu}\, d t^{\mu} \right], \quad \hat{H}_0\equiv \hat{H}_{\rm BCS}.
\label{path1}
\ee
$\hat{U}$ does not depend on the  path $\cal{P}$, except its initial and final  points. This invariance follows from the fact that  the
  gauge field with components ${\cal {A}}_{\mu}=-i\hat{H}_{\mu}$  has zero curvature. Hence, its integral over any closed path that does not enclose singularities of $\hat{H}_{\mu}$ is zero  \cite{sinitsyn2017integrable}.

\begin{figure} [t]
{\includegraphics[width=0.9\columnwidth]{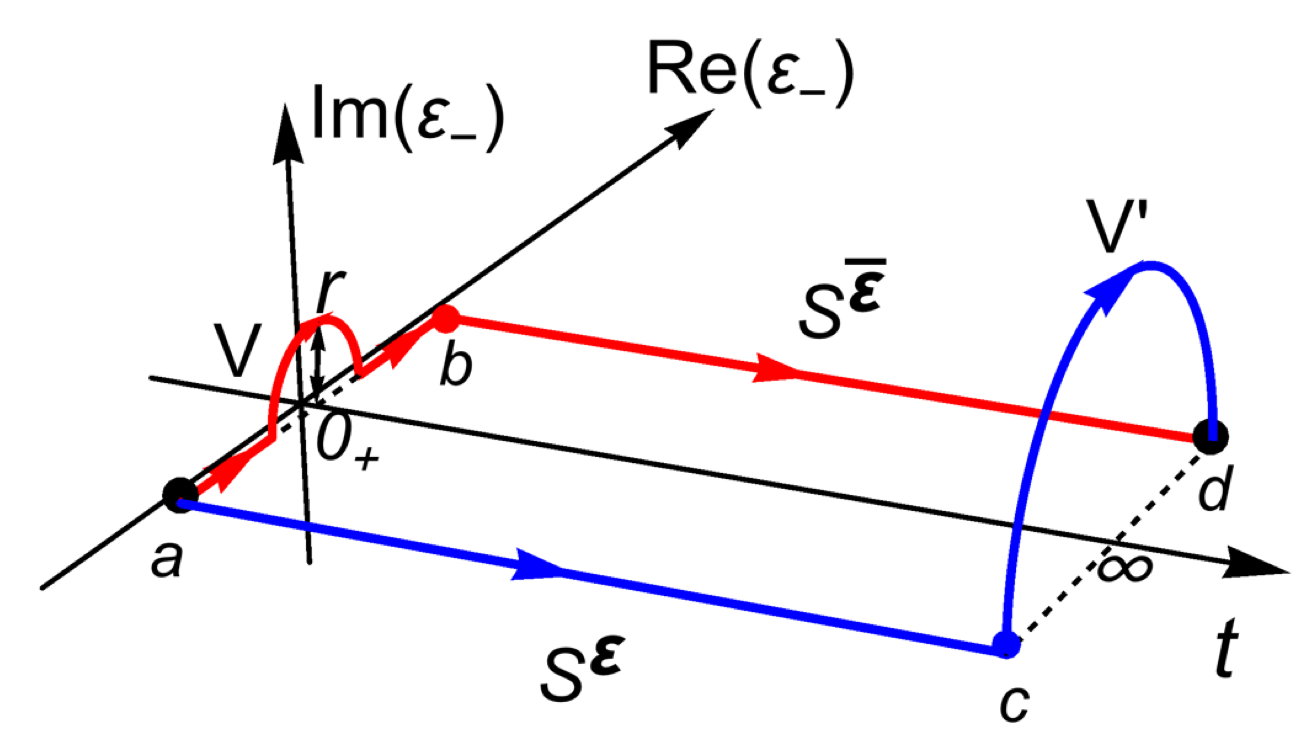}}
\caption{{ Two  paths corresponding to the same evolution operator}. Evolution takes place over the space of real time $t$ and complex values of  $\ve_-\equiv \ve_j-\ve_{j+1}$. The initial point $a$ corresponds to $t=0_+$ and $\ve_j< \ve_{j+1}$. The final point $d$ is at $t\rar \infty$ and $\bar{\ve}_{j}=\ve_{j+1}$, $\bar{\ve}_{j+1}=\ve_{j}$. The red path $a\rar b \rar d$ avoids the singularity at $\ve_{-}=0$ from the infinitesimally small distance $r$ at $t=0_+$, and  the blue path $a\rar c \rar d$ avoids this singularity at $t\rar \infty$ along the arc $(cd)$ with a finite radius. Evolution over  links $(ab)$, $(bd)$, $(ac)$, and $(cd)$ is described by  matrices, respectively, $V$, $S^{\bar{\bm \ve}}$, $S^{\bm \ve}$, and $V'$.
}
 \label{fig:contour}
\end{figure}
  Let us  compare two evolution paths shown in Fig.~\ref{fig:contour} that start
at vector ${\bm \ve}$ at $t=0_+$  (point $a$) and end at $t\rar \infty$ at vector $\bar{\bm \ve}$  (point $d$) such that two adjacent in magnitude vector components are related by  $\bar{\ve}_j = \ve_{j+1}$ and $\bar{\ve}_{j+1}= {\ve}_j$ for one and only one $j$, while in all other respects  ${\bm \ve}$ and $\bar{\bm \ve}$ are identical.   These paths have to avoid the singularity of $\ \hat{H}_j-\hat{H}_{j+1}$ at $\ve_j=\ve_{j+1}$, so the difference $\ve_{-}\equiv \ve_j-\ve_{j+1}$ is allowed to be complex valued.

At the  path $a\rar b \rar d$, evolution  matrix $V$ along the link $(ab)$  reverses the sign of $\ve_{-}$ keeping  other parameters constant.  Next, at $(bd)$, we keep  $\bar{\bm \ve}$ constant and evolve to the end point  at $t\rar \infty$ with the evolution matrix $S^{\bar{\bm \ve}}$. At the other path $a\rar c \rar d$ we initially evolve, with the evolution matrix $S^{{\bm \ve}}$, along the real time to a point at large  $t$ and  then reach the end point, with the evolution matrix $V'$, at constant $ t$.
The invariance of $\hat{U}$ means that
\be
S^{\bar{\bm \ve}} V|\psi_0 \ra= V'S^{{\bm \ve}} |\psi_0 \ra.
\label{int-cond1}
\ee
We will use Eq.~(\ref{int-cond1}) to compare amplitudes of evolution along real $t$ from $|\psi_0 \ra$ to states   $| j \ra = | \ldots, \ua_j, \da_{j+1}, \ldots \ra$ and $|\tilde{j} \ra = |\ldots, \da_j, \ua_{j+1}, \ldots \ra$  that are different only by directions of  two  spins with neighboring  $\ve_j$ and $\ve_{j+1}$.

Consider first the link $(ab)$ in Fig.~\ref{fig:contour}.
Suppose that initially $\ve_j<\ve_{j+1}$. We keep $\ve_j+\ve_{j+1}$ constant, so
\be
\nonumber \int H_j \, d \ve_j + \int H_{j+1} \,d \ve_{j+1} = (1/2) \int (\hat{H}_j -\hat{H}_{j+1}) \,d\ve_{-} .
\ee
The evolution operator for this link  is
\be
{V}=\hat{\cal T} {\rm exp} \left[-(i/2) \int_{{\cal{P}}_{(ab)}}  (\hat{H}_j -\hat{H}_{j+1}) \,d \ve_{-} \right].
\label{ev1}
\ee
All  $\hat{H}_{\mu}$ commute, so $|\psi_0\ra$ is the eigenstate of not only $\hat{H}_{\rm BCS}$ but also of $\hat{H}_j -\hat{H}_{j+1}$. Hence, $\la \alpha | (\hat{H}_j -\hat{H}_{j+1}) |\psi_0 \ra =0$ for $|\alpha  \ra \perp |\psi_0 \ra$. We calculate $|\la \psi_0|V| \psi_0 \ra|$ bypassing the singularity at $\ve_j=\ve_{j+1}$ along the semicircle  of radius $r$ in the complex  $\ve_{-}$ plane. Only the piece of this path with nonzero ${\rm Im}(\ve_{-})$  contributes to the absolute value.
In the limit $r\rar 0$ at $t=0$, we have
$
\hat{H}_j -\hat{H}_{j+1} \rar - 4g\hat{\bm s}_j\cdot \hat{\bm s}_{j+1}/\ve_{-}.
$
For $\ve_{-}=re^{i\phi}$, we find
\be
|\la \psi_0|{V}|\psi_0\ra| \!=\! e^{ -2g \int_{\pi}^{0} d \phi \,   \la \psi_0| \hat{\bm s}_j\cdot \hat{\bm s}_{j+1} |\psi_{0} \ra_{t=0} }   \!=\! e^{\pi g/2}.
\label{v00}
\ee

Consider now the link $(cd)$, at which $t \rar \infty$.
If $n \ne j,j+1$ we have $\hat{H}_n =t s^z_n+O(1)$. Hence, such   Hamiltonians are proportional to spin operators, and commutation of $\hat{H}_n$ with $\hat{H}_j-\hat{H}_{j+1}$
means conservation of  $s_n^z$ during the evolution along this link, i.e., $\la j| V'| \alpha \ra=0$ if $|\alpha \ra$ has different from $|j\ra$  value  of a spin with index $n$.  Transitions
between states $|j\ra$ and $|\tilde{j}\ra$, however, should be treated with extra care
because $\hat{H}_j$ and $\hat{H}_{j+1}$ are singular near
$\ve_j=\ve_{j+1}$ where conservation of spins with indexes $j$ and $j+1$ breaks down.  So,  we set evolution between points $c$ and $d$ along a semicircle with a finite radius in Fig.~\ref{fig:contour},  restricting this evolution  to the subspace of states  $|j\ra$ and $|\tilde{j} \ra$.

Let us again change variables so that $\ve_{-}=b s /t$, where $b/t \rar 0$ and $b>0$ is finite. The
large parameter $t$ then drops out of the evolution equation along $(cd)$:
\be
i\frac{d |\psi  \ra}{ds}  = \left(
\begin{array}{cc}
b+ g/(2s) & \kappa/s \\
\kappa/s & -b + g/(2s)
 \end{array}
 \right) |\psi  \ra ,
\label{shr1}
\ee
where $|\psi \ra = c_j(t) |j\ra +c_{\tilde{j}}(t) |\tilde{j} \ra$ with amplitudes $c_j$ and $c_{\tilde{j}}$; $s$ changes along a semicircle $s=Re^{i\phi}$ with $R\rar \infty$, and $\phi$ decreases from $\pi$ to $0$. Parameter $\kappa$ is a constant that depends on states of all spin directions in $|j\ra$.
In (\ref{shr1}), we dropped all terms that decrease faster than $\sim 1/R$.

This evolution was already studied in Ref.~\cite{sinitsyn-14pra}, according to which we can disregard the vanishingly small off-diagonal terms $\kappa/s$ in calculation of the diagonal elements of $V'$:
\be
|\la j| V'| j  \ra|= e^{ - 2g \int_{\pi}^{0} d \phi \,   \la j| \hat{\bm s}_j\cdot \hat{\bm s}_{j+1} |j \ra }   = e^{-\pi g/2}.
\label{vjj}
\ee
As for the off-diagonal elements of $V'$, such an adiabatic approximation can be justified  only if the initial state has the lower energy at $s \rar -\infty$. Only then cannot the evolution along the complex time contour lead to growth of the inter-level transition amplitude \cite{sinitsyn-14pra}. For $\ve_j<\ve_{j+1}$ this means that
\be
\la j| V'| \tilde{j}  \ra=0, \quad \ve_j <\ve_{j+1},
\label{vjjp}
\ee
 independently of $\kappa$ but we generally have $\la \tilde{j}| V'| j \ra\ne 0$.
The latter element does not appear in the following calculations but we note that such a nonzero term
would be relevant if the singularities were enclosed by the paths with ${\rm Im}(\ve_{-})<0$  instead of those in Fig.~\ref{fig:contour}.

\begin{figure} [t]
{\includegraphics[width=0.95\columnwidth]{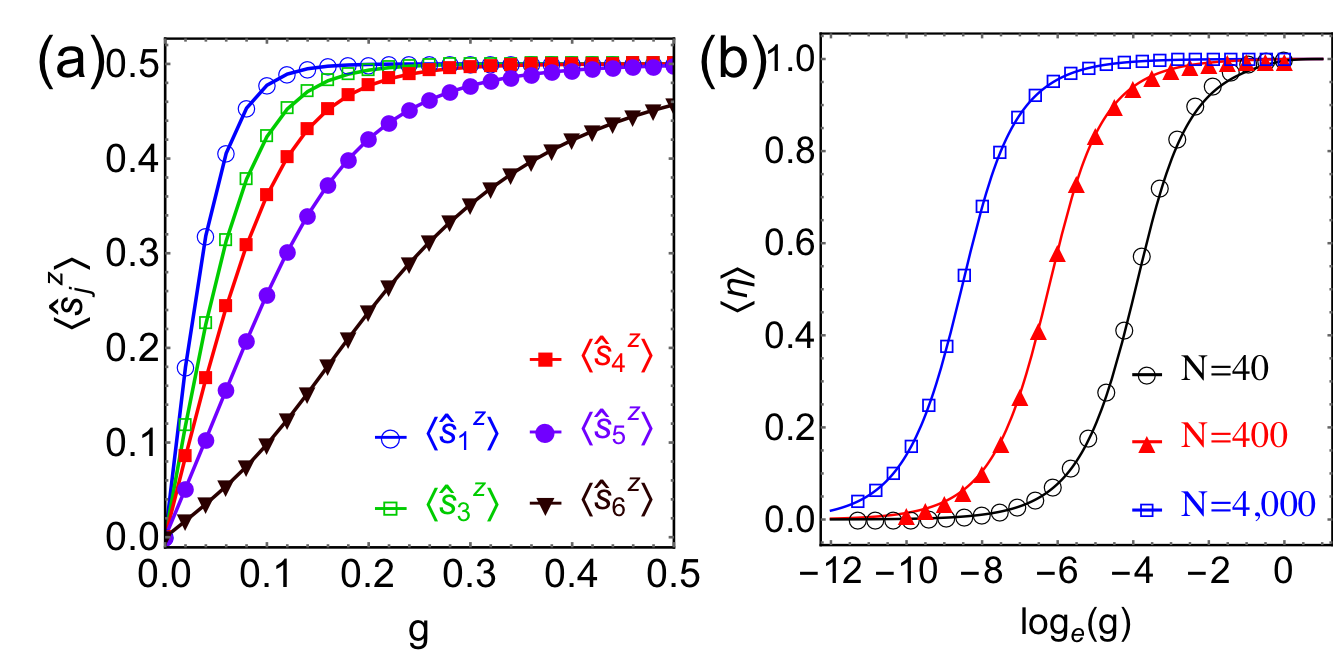}}
\caption{
{\bf (a)} The final polarization of several spins for $N=12$ and $S_{\rm tot}^z=0$. Prediction of the Gibbs distribution (solid curves) is compared to numerical solution of the Schr\"odinger equation (point markers) \cite{supp}.  Here,  $\ve_j$  are the same as in Fig.~1.
   {\bf (b)} Accuracy of QA  at different $g$ and $N$ at $t \rar \infty$.  Points show exact predictions of Eq.~(\ref{eq:Psjf}), and solid lines are the large-$N$ approximation (\ref{eq:M2}).}
 \label{Man}
\end{figure}
Evolution along  $t$ at constant  $\bar{\bm \ve}$ is the same as at ${\bm \ve}$ but with exchanged spin indexes: $j\leftrightarrow j+1$.
So, $\la j|S^{\bar{\bm \ve}}| \psi_0 \ra=\la \tilde{j} |S^{{\bm \ve}}|\psi_0 \ra$. The  probabilities to find the microstates $|j\ra$, $|\tilde{j}\ra$  at fixed ${\bm \ve}$ and $t\rar \infty$ are then, $P_{|j\ra}=|\la j|S^{{\bm \ve}}|\psi_0 \ra|^2$ and $P_{|\tilde{j}\ra } =|\la j|S^{\bar{\bm \ve}}|\psi_0 \ra|^2$.
Multiplying both sides of equation~(\ref{int-cond1}) by $\la j |$ from the left, and using (\ref{v00}), (\ref{vjj}), and (\ref{vjjp}), we find that transition probabilities from $|\psi_0\ra$ to the two states   are related:
\be
P_{|\tilde{j}\ra}/P_{|{j}\ra} =  e^{-2\pi g}, \quad \ve_j<\ve_{j+1}.
\label{ratio}
\ee
Equation~(\ref{ratio}) is valid for any index $j$ and arbitrary values of all parameters $\ve_k \notin (\ve_j,\ve_{j+1})$ and spin projections $s^z_k$ in $|j\ra$ for $k\ne j, j+1$.
{It has the form of the detailed balance condition that is possible to satisfy only if the probability to find any final eigenstate of $\hat{H}_A$,
$|\{ s^z \} \ra \equiv |s^z_1, s^z_2, \ldots, s^z_{N} \ra$, is given by the Gibbs distribution }
\be
P_{\{s^z\}} = \frac{1}{\cal Z} e^{- 2\pi g \sum_{j=1}^{N} j s^z_j} \delta \left( \sum_{j=1}^{N} s^z_j - S^z_{\rm tot} \right) ,
  \label{eq:Psjf}
  \ee
where  $1/{\cal Z}$ is a normalizing factor.
In Fig.~\ref{Man}(a), we test Eq.~(\ref{eq:Psjf}) numerically and illustrate that  generally spins aline  along their ground state directions at  $g\ll 1$, i.e., in strongly nonadiabatic regime. { Independence of $P_{\{s^z\}}$ of $\ve_j$, except the relative order of these parameters, is the  property shared by many solvable time-dependent models for reasons discussed in \cite{sinitsyn-jpa18}.}   A simpler proof of this independence is via relation (\ref{int-cond1}) applied to a situation with ${\bm \varepsilon}$ and $\bar{\bm \varepsilon}$ different only by continuous changes that keep all vector components real and nondegenerate. Pieces of evolution in Fig.~\ref{fig:contour} with $t={\rm const}$ then do not contribute to transition probabilities at all, and (\ref{int-cond1}) leads to relation $|S^{{\bm \ve}}|=|S^{\bar{\bm \ve}}|$.

{The Gibbs distribution may not  describe the thermalized state of the right Hamiltonian. However, for equidistant spin splittings, $\varepsilon_j=\ve j$,  the distribution~(\ref{eq:Psjf}) does correspond to  $\hat{H}_{\rm BCS}$ at $\rm t \rar \infty$, i.e. we find a complete thermalization in this case,} as we announced in (ii), at temperature
\be
\nonumber T=\ve/(2 \pi k_B g),
\label{temp}
\ee
where $k_B$ is the Boltzmann constant.

To derive coarse-grained characteristics at $N \gg 1$, it is safe to replace the delta-function in (\ref{eq:Psjf}) by a weaker constraint that equates only the average spin to $S_{\rm tot}^z$ (see supplementary material
 \cite{supp} for details of calculations, which includes Refs.~\cite{Pathira,LL-QM-3}).  This leads to
\be
 \la \eta \ra \approx \frac{2}{\pi g N} ( \log (1+e^{\pi g N}) -\log2 )  - 1,
 \label{eq:M2}
\ee
which we confirm in Fig.~\ref{Man}(b), and from which we find that to achieve accuracy $\la \eta\ra$ at conditions $S_{\rm tot}^z=0$, $N\gg1$, $g\gg 1/N$,  we should set $g=2\log 2/[\pi N (1-\la \eta \ra)]$ that is
far from the adiabatic regime at $N \rar \infty$,  proving  (i).

 For example, if  $g=0.01$, i.e., calculations are $100$ times faster than the adiabatically protected ones,
the probability of a wrong result per spin is $(1-\la \eta \ra)/2 \approx 22./N$, for $N\gg1$, and only $20$-$25$ errors appear totally in the limit $N\rar \infty$. We note that experiments with the BCS Hamiltonian  in ultracold atoms deal with $N\sim10^6$ fermions \cite{bcs-atoms}, in which BCS coupling can be controlled by time-dependent fields. 

 Our solution illustrates importance of quantum correlations that are introduced by $\hat{H}_B$: collective effects help some of the spins to settle much earlier in time (Fig.~\ref{Man}(a)). The remaining spins in turn feel this, which helps them to find their own ground state directions faster while satisfying the total spin conservation constraint.
 If, otherwise, we  had set $\hat{H}_B=\sum_{i=1}^N \hat{s}_i^x$, i.e., if we were looking for the ground state of permanently uncoupled spins, we would find the final $\la \eta \ra$  independent of $N$ and decaying quickly at $g<1$, independently of the choice of $\ve_j$.

This proves that strongly interacting  QA dynamics can be studied exactly beyond the models of noninteracting fermions and their equivalents \cite{Dziarmaga2005}.
Unlike these models, simplicity of the final distribution (\ref{eq:Psjf})  rather reflects the facts that  $g(t) \sim 1/t$ is scale-free and the model~(\ref{h-bcs0}) likely has no conservation laws, except $S^z_{\rm tot}={\rm const}$.
The latter difference leads to essentially different behavior of error probabilities in the nonadiabatic regime. Thus, the QA models that are equivalent to sets of independent two-level systems, such as the quantum Ising chain in a transverse magnetic field \cite{Dziarmaga2005}, inevitably predict the linear scaling of the number of computational errors with growing $N$  at other conditions fixed. In contrast, our model shows a vanishing error probability {\it per spin} in the limit of large $N$  in the nonadiabatic regime at a fixed driving protocol and spin coupling distribution. This observation suggests that QA protocols with a strongly entangled initial state may provide considerable boost to accuracy of QA computations. Further experimental and numerical evidence in support of this conclusion is still needed to understand advantages of this approach.

 Quantum thermalization  is usually associated with semiclassical chaos that makes  local operator expectations in typical eigenstates close to thermal ensemble averages  \cite{kinoshita2006quantum,rigol2008thermalization}.  We showed, however, that also regular fields can steer coherent reversible evolution toward  the perfect Gibbs  distribution of all independent eigenstates of a Hamiltonian.
Emergence of the strong detailed balance constraint, which enables this thermalization, would be impossible without the symmetry responsible for model's integrability. So,  integrability is not only compatible but it is  needed naturally to find the Gibbs distribution after QA.
 In \cite{supp}, we support this conclusion by showing  that the model's  symmetry reflects invariance of the evolution matrix under action of the Braid Group and the associated with it quantum group ${\rm SU}_{q}(2)$ \cite{D-85,D-86,CGO-92} where the deformation parameter $ q\equiv e^{-\ve /2k_BT } $ defines the  temperature scale.


 Existence of such a QA path to complete thermalization may have fundamental consequences.  Thus, the Universe can be a closed many-body quantum system that has passed through inflation with  changing  parameters and early entanglement \cite{cosmology}, just like during QA. Everything now seems going toward the globally thermalized state but; for a closed system with many visible symmetries, is this  expected? Property (ii) means that this is possible, and if such thermalization is  realized in our world, then (iii) means that the most fundamental equations of physics are integrable, which in turn provides a hint as to why mathematics is efficient in describing the natural laws \cite{wigner1995unreasonable}.

{\it Acknowledgements.} This work was carried out under the auspices of the National Nuclear Security Administration of the U.S. Department of Energy at Los Alamos National Laboratory under Contract No. DE-AC52-06NA25396 (N.A.S. and F. Li).  V.Y.S. was supported by NSF Grant No. CHE-1111350 (V.Y.C.). F. Li and N.A.S. also thank the support from the LDRD program at LANL. F. Li is also supported by the Fundamental Research Funds for the Central Universities from China. 

Authors equally contributed to this article.



%



\begin{widetext}
\section*{Supplementary Material for ``Quantum annealing and thermalization: insights from integrability"}
\end{widetext}
\subsection*{Characteristics of the Gibbs distribution}
Here, we explore the Gibbs distribution in Eq.~(12) of the main text and provide details of derivation of Eq.~(13) and ground state probability, plotted in Fig.~2(a) there. 

\subsubsection*{Ground state probability}

First, we note that the transition probability to an arbitrary microstate $|\{ s^z \} \ra \equiv |s^z_1, s^z_2, \ldots, s^z_{N} \ra$ can be written explicitly:
\be
\label{tr-prob1}
P_{\{ s^z \}} = \prod_{j=1}^{N} p^{\sigma_j}_{m_j, j}, 
\ee
where $\sigma_j=\pm$ for, respectively, $s_z^j=\pm 1/2$;  $m_j =2(S^z_{\rm tot}-\sum_{l=j+1}^{N} s^z_l)$, and where
\be
\nonumber p^{-}_{m, n} =\frac{1-x^{(n-m)/2}}{1-x^{n}}, \quad
p^+_{m, n} = 1- p^{-}_{m, n}, \quad x \equiv e^{-2\pi g}. \label{pmn}
\ee
To verify that (\ref{tr-prob1}) is the same as the Gibbs distribution in Eq.~(12) of the main text, one can take the ratio of probabilities of any two states with flipped spins that have nearby indexes. The result coincides with Eq.~(11) in the main text that has  lead to the Gibbs distribution.

From (\ref{tr-prob1}), the  probability to remain in the ground state at $t\rar \infty$ when $S_{\rm tot}^z=0$ is
$P_G = (x,x)_{N/2}/(x^{N/2+1},x)_{N/2}$,
where $(a,q)_k \equiv \prod_{i=0}^{k-1}(1-aq^i)$ is the q-Pochhamer symbol.
At large $N$, this probability is independent of $N$: 
\be
P_{G}^{N \rar \infty} = (x,x)_{\infty}.
\label{prob-24}
\ee

\subsubsection*{Coarse-grained characteristics}
Next, to derive the mean  number of errors at $N\gg1$,  we
recall a well known duality between the spin BCS Hamiltonian (3) in the main text and interacting fermions \cite{yuzbashyan2015quantum}.  In fermionic representation, at $t\rar \infty$, the spin ground state at $S^z_{\rm tot}=0$ corresponds to the Fermi sea  of $N/2$ noninteracting fermions filling the lowest half of $N$ energy levels.    In the thermodynamic limit $N \gg 1$, it is safe to  approximate  the canonical distribution  by the  grand canonical one for calculation of basic statistical characteristics of the noninteracting Fermi gas \cite{Pathira}.  Returning to the spin language, this means that we can replace the constraint due to the delta-function in Eq. (12) in the main text by the chemical potential $\mu$ that fixes only the value of the average spin: 
\be
P_{\{s^z \}, \mu} = \frac{1}{\cal{Z}} e^{-\beta \sum_{j=1}^N (j-\mu) s_j^z  } .
\ee
The average polarization of each spin in this approximation is $\la s^z_j \ra = \frac{1}{2}\tanh [\beta (\mu-j)/2] $.  To guarantee that $\la S_{\rm tot}^z\ra=0$, we should set $\mu = (N+1)/2 $. Then $\la \eta \ra = \frac{2}{N} \sum_{j=1}^{N/2} \tanh [\beta (\mu-j)/2] $. Taking the continuous limit, we convert this sum into integral, leading to
\be
 \la \eta \ra \approx \frac{2}{\pi g N} ( \log (1+e^{\pi g N}) -\log2 )  - 1,
 \label{eq:M221}
\ee
which we compare with exact predictions of the Gibbs distribution in Fig.~4(b) in the main text.
For $1>g  \gg 1/N$, equation~(\ref{eq:M221}) simplifies to $\la \eta \ra \approx 1-2\log 2/(\pi g N)$, which can be inverted to obtain the estimate of $g$  that guarantees precision $\la \eta \ra$.

\subsection{Entropy of the excitation distribution}
Apart from $\eta$, another measure of QA precision is the  entropy of the final distribution:
 \be
S=-\sum_{\{ s^z\}} P_{\{ s^z \}} \log P_{\{ s^z \}}, \label{eq:sdef}
\ee
where  summation runs over all the microstates of the Ising spin Hamiltonian.

At $g\rar 0$, the final state coincides with the fully entangled initial state. This leads to equiprobable microstates of $\hat{H}_A$ (infinite temperature).
 Since the size of the Hilbert space for $N$ spins with $S^z_{\rm tot}=0$ is given by $N_h = C_{N}^{N/2}$, this case corresponds to $S(g\rar 0) = \log N_h $. Using the Sterling's approximation, in the large $N$ limit we find then $S\sim N \ln 2$, i.e., entropy is growing linearly with $N$ in the limit of fast QA.

 \begin{figure} [!htb]
{\includegraphics[width=0.95\columnwidth]{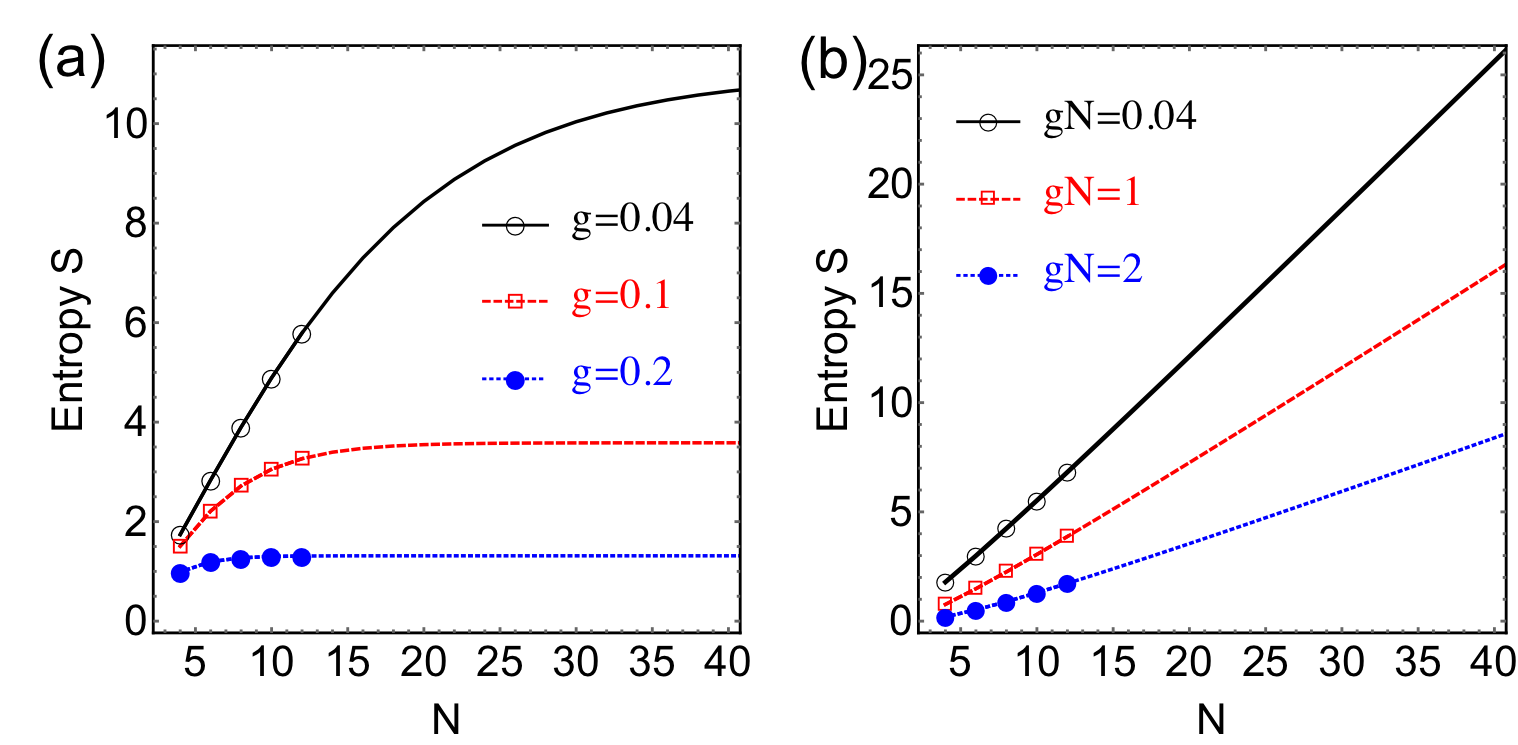}}
\caption{{\bf (a)} Entropy as a function of $N$ for different coupling constants $g=0.04, 0.1, 0.2$. {\bf (b)} Entropy as a function of $N$ for different fixed $gN$. The points denoted by plot markers are produced by exact numerical calculation, and the solid lines are calculated by using Eq.~(\ref{eq:S}). } \label{Fig:entropy}
\end{figure}
 For finite $g$, the entropy can be found from the partition function ${\cal Z}$:
\be
S= {\rm log}{\cal Z} - g\frac{\partial {\rm log} {\cal Z}}{\partial g}. \label{eq:S}
\ee
Having explicit expression for the ground state probability, the partition function of the Gibbs distribution can be obtained for $S_{\rm tot}^z=0$ as ${\cal Z} = x^{N^2/8} P_{G}^{-1}$.
 For  fixed $g$ and
$N \gg 1/g$, we take the continuous limit and find that $S$ saturates at
\be
S\sim \frac{1}{\pi g} Li_2(e^{-\pi g}) - \frac{1}{2} \log(1-e^{-\pi g}).
\ee
where $Li_2(z) \equiv \sum_{j=1}^{\infty} z^j/ j^2$ is the 2nd order polylogarithm function whose leading orders are $Li_2(z) = z + z^2/4 + \ldots$.
Figure~\ref{Fig:entropy} compares Eq.~(\ref{eq:S}) with numerical results that were obtained by direct calculations of all microstate probabilities and then using Eq.~(\ref{eq:sdef}). Figure~\ref{Fig:entropy}(a)
shows that total entropy of the final distribution saturates at a finite value in the limit $N\rar \infty$, which is consistent with our result in the main text that the number of errors is restricted to $O(N^0)$ values in this limit.

Fixing the product $gN={\rm const}$, $S$ scales linearly with $N$, as shown in Fig.~\ref{Fig:entropy}(b). Hence, the thermodynamic limit at $N\rar \infty$ is well defined by keeping $gN$ constant. I order to fix the error probability per spin then one should adjust $g$ to be $g\sim 1/N$. Since computation time is roughly proportional to $g$, this means that we should shift the rate of annealing further from the adiabatic regime.

\subsection*{Numerical calculation of $\la \eta \ra$ and $P(\eta)$ at  large $N$}

 Since dimension of the Hilbert space of $N$ spins-$1/2$ with $S^z_{\rm tot}=0$ is exponentially increasing with $N$,
numerically exact calculations of average characteristics become problematic even using explicit formulas for final probabilities of microstates. 
 Here we provide a method that we used to calculate the distribution function $P(\eta)$ of $\eta$ during time that scales as $N^2$ with the number of spins.
 We used this method to generate Fig.4(b) in the main text.
 \begin{figure} [!htb]
{\includegraphics[width=0.75\columnwidth]{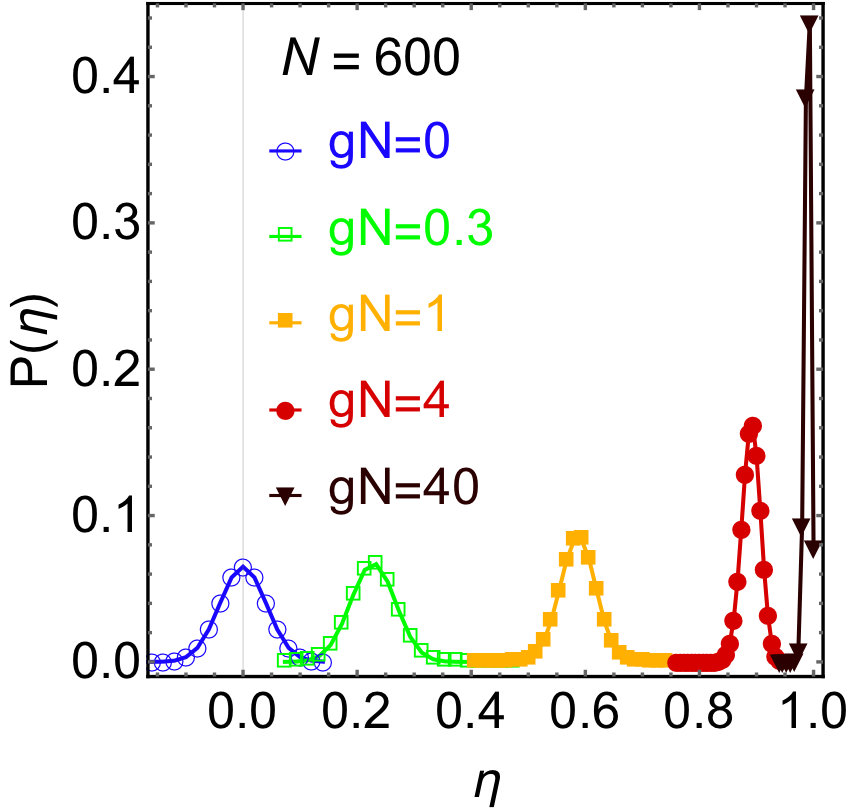}}
\caption{The probability distribution of  accuracy $\eta$ of QA for different couplings $g$ and $N=600$.  Plot marker points are  numerically exact predictions of the  distribution in Eq.~(12) of the main text, and solid lines are the approximation by a Gauss curve with variance (\ref{var1}) and the mean (\ref{eq:M221}). } \label{Fig:peta}
\end{figure}
 
 Let  $\ve_1 < \ve_2 < \ldots <\ve_{N}$.
The explicit formula (\ref{tr-prob1}) for the probability of any microstate  has such a structure  that this  probability  can be calculated by determining the state probability of each spin sequentially, starting from the $N$-th spin and continuing down to the spin with index $1$. For the $j$-th spin, let us define a probability $q_{j, m}$ in which $m$ is the  difference between numbers of  spins up and  spins down with larger than $j$ indexes.  Probability
$q_{j, m}$ depends only on such probabilities of the $(j+1)$-th spin: $q_{j+1, m-1}$ if the $j$-th spin is finally up, and $q_{j+1, m+1}$, if the $j$-th spin is finally down.  This leads to a Markov chain equation: 
\be
q_{j, m} = p^{+}_{m_j, n_j}  q_{j+1, m-1} + p^{-}_{m_j, n_j} q_{j+1, m+1},
\ee
where $p^{\pm}_{m_j, n_j}$ are given by (\ref{pmn}). We solved this equation numerically recursively.
After the $(N/2+1)$-th step, we determined $q_{N/2, -m}$ which is the probability distribution of the final polarization of the half of spins with the lowest indexes, from which we obtained 
the distribution of accuracy, $P(\eta)$, by identifying $\eta=-2m/N$. 

We used this algorithm to find $P(\eta)$ in  systems with up to 4000 spins.  Comparison between this numerical calculation and analytical calculation of $\la \eta \ra$ has been demonstrated in Fig.~4(b) of the main text. Repeated QA leads also to fluctuations of $\eta$ from one calculation to another. At $N\gg1$, $\eta$ has the Gaussian distribution, which is  
 peaked near the average value. In the grand canonical ensemble, $\la s^z_j s^z_k \ra = \la s^z_j \ra \la  s^z_k \ra $ for $j \neq k$, so we can estimate the variance of this distribution:
\be
{\rm var}({\eta}) \equiv \la\eta^2 \ra - \la \eta \ra^2 \approx \frac{4}{\pi gN^2} \Big(\frac{1}{1+ e^{-\pi g N }} -\frac{1}{2} \Big).
\label{var1}
\ee
We  compare such a Gaussian approximation of  $P(\eta)$ with results of numerical calculations of $P(\eta)$ in Fig.~\ref{Fig:peta}.

\subsection*{Interchange Symmetry, Monodromy, and Quantum Groups}
\label{sec:BCS-S-symm-QG}

In the main text, we  identified the structure of the final state at $t \to \infty$ by combining the fact that interchange of any two spins commutes with BCS evolution with a  simple topological argument that implies, due to the zero-curvature condition, that the evolution in the multi-time space depends on the topological (homotopy) type of the integration path. Here, we rationalize that this argument is the implementation of symmetry that naturally leads to  the quantum group ${\rm SU}_{q}(2)$, which is the real ``compact'' version of a complex quantum group ${\rm SL}_{q}(2; \mathbb{C})$, with $q = e^{-\pi g}$.

Invariance of the evolution operator (Eq.~(5) of the main text) can be presented as
\begin{eqnarray}
\label{suppl-commute} \hat{U}(p_{j, j+1}(l_{ca})) \hat{U}(l_{ ba}) = \hat{U}(l_{ dc}) \hat{U}(l_{ ca}),
\end{eqnarray}
where $\hat{U}$ is the evolution operator defined in the main text, points $a,b,c,d$ are defined in Fig.~3 there and $p_{jk}$ denotes the permutation map in the parameter space that interchanges $\varepsilon_{j}$ with $\varepsilon_{k}$, and $l_{\beta\alpha}$ is the path that connects $\alpha$ to $\beta$,  in Fig.~3; note that we have used $l_{ db} = p_{j, j+1}(l_{ca})$. Let us also define $\hat{p}_{jk} = 2 (\hat{\bm{s}}_{j} \cdot \hat{\bm{s}}_{k}) + (1/2)$ for a permutation operator in the spin space. We further make use of the  symmetry of our system that interchange of any two energies, described by $p_{jk}$ maps, accompanied with interchange of the corresponding spins, described by $\hat{p}_{jk}$ operators, does not change the equations. This implies $\hat{p}_{j, j+1} \hat{U}(p_{j, j+1}(l_{ ca})) \hat{p}_{j, j+1} = \hat{U}(l_{ ca})$, allowing Eq.~(\ref{suppl-commute}) to be recast in a form
\begin{eqnarray}
\label{suppl-commute-2} \hat{U}(l_{ca}) \hat{\sigma}(l_{ba}) = \hat{\sigma}(l_{dc}) \hat{U}(l_{ca}),
\end{eqnarray}
where we have introduced the monodromy operators/matrices $\hat{\sigma}(l_{ ba}) = \hat{p}_{j, j+1} \hat{U}(l_{ ba})$ and $\hat{\sigma}(l_{ dc}) = \hat{p}_{j, j+1} \hat{U}(l_{ dc})$, associated with the paths $l_{ba}$ and $l_{dc}$ that connect $a$ to $b = p_{j, j+1}(a)$ and $c$ to $d = p_{j, j+1}(c)$, respectively, so that Eq.~(\ref{suppl-commute-2}) means that monodromy commutes with evolution. It is important to note that usually monodromy is associated with closed paths/loops; our situation is reduced to the standard one by making use of the particle interchange symmetry and introducing the so-called configuration space by announcing the points in the parameter space, which differ just by a permutation of the energies $\varepsilon_j$, identical. In the configuration space the paths $l_{ ba}$ and $l_{ dc}$ become loops, while $l_{ db} = p_{j, j+1}(l_{ ca}) = l_{ ca}$, and we recover the standard monodromy setting.

Loops in the configuration space are naturally represented by braids with the path $l_{ ba}$ that interchanges $\varepsilon_{j}$ with $\varepsilon_{j+1}$, usually denoted $\sigma_{j}$, being illustrated in Fig.~\ref{braid-fig}(a). The braids can be multiplied using concatenation, so that the braids form a group, generated by the elementary braids $\sigma_{k}$ with $N$ strands, denoted by ${\rm B}_{N}$. Since the braids represent homotopy classes of paths, the BG has relations $\sigma_{j} \sigma_{j+1} \sigma_{j} = \sigma_{j+1} \sigma_{j} \sigma_{j+1}$ (cubic relations, illustrated in Fig.~\ref{braid-fig}(b).) and $\sigma_{j} \sigma_{k} = \sigma_{k} \sigma_{j}$ for $|j-k| \le 2$ (obvious relations). Associating the monodromy matrix $\hat{\sigma}$ with any braid $\sigma$ by
\begin{eqnarray}
\label{M-gen-braid} \hat{\sigma} = \hat{p}(\sigma) \hat{U}(\sigma),
\end{eqnarray}
with $p(\sigma)$ being the permutation, associated with the braid $\sigma$, which builds a $2^{N}$-dimensional representation of ${\rm B}_{N}$. The BG commutes with dynamics, in particular, considering dynamics from $t = 0_+$ to $t = \infty$, we have $S \hat{\sigma}_{ a} = \hat{\sigma}_{ c} S$ for the scattering matrix $S$ that connects the correlated states at $t \to 0$ to their  counterparts at $t \to \infty$, and any braid $\sigma$, with $\sigma_{a}$ and $\sigma_{c}$, representing the same braid, defined with respect to the base points ${ a}$ and ${ c}$ in Fig.~3 of the main text, located at $t \to 0_+$ and $t \to \infty$, respectively.

\begin{figure} [!htb]
{\includegraphics[width=0.96\columnwidth]{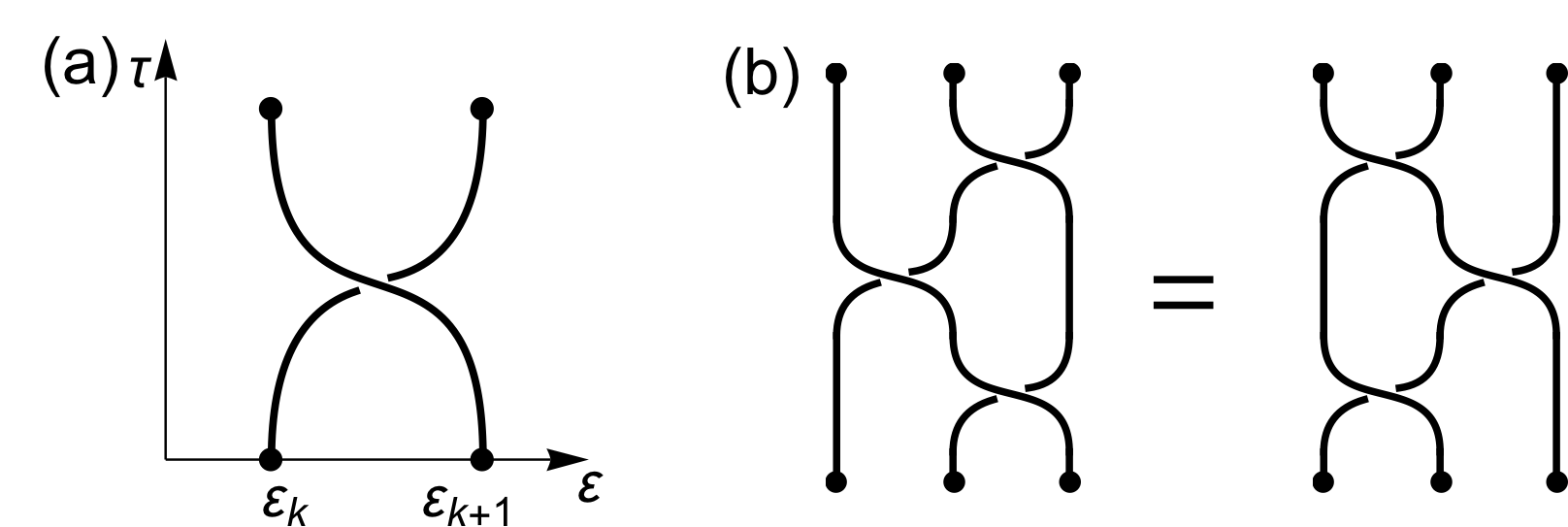}}
\caption{{\bf (a)} The simplest element of the braid group that exchanges nearby indexes of two spins during evolution in multi-time space parametrized by $\tau$. {\bf (b)} Symmetry of the braid group represented by Yang-Baxter-Zamolodchikov relation for R-matrix  in Eq.~(\ref{supplYBZ-equation}). } 
\label{braid-fig}
\end{figure}

A quantum group appears  in considering the monodromy using the language of  states in the $t \to \infty$ limit. Indeed, adiabatic states in the  $t \to \infty$ region are well-defined, and in computing the monodromy $\hat{\sigma}_{j}$ we can keep the pair $\varepsilon_{j}, \varepsilon_{j+1}$ of energies well-separated from the rest. So, the monodromy involves only the $j$-th and $(j+1)$-th spins, and does not depend on $j$. This means that $\hat{\sigma}_{j} = \sigma_{j, j+1}$ for some $4 \times 4$ matrix $\sigma$ that can be  computed  explicitly by considering a $4 \times 4$ scattering problem for two elementary spins with indexes $j$ and $j+1$. Due to conservation of the total spin, this matrix factorizes into two scalar problems and the $2 \times 2$ one of a confluent hypergeometric type, as described in the main text for elements of the matrix $V'$, but even this is not necessary. By representing $\sigma = \hat{p} R$, with $\hat{p}$ being just the spin permutation operator, the cubic relations of the BG adopt a form
\begin{eqnarray}
\label{supplYBZ-equation} R_{j, j+1} R_{j, j+2} R_{j+1, j+2} = R_{j+1, j+2} R_{j, j+2} R_{j, j+1}
\end{eqnarray}
of the  Yang-Baxter-Zamolodchikov (YBZ) or triangle equation, whose solutions have been classified, and in the $2 \times 2$ case of interest are represented by a one-parameter family 
\begin{widetext}
\begin{eqnarray}
\label{suppl-R-2x2} R = (1/\sqrt{q})(I \otimes I + (q-1) (X_{11} \otimes X_{11} + X_{22} \otimes X_{22}) + (q - 1/q) X_{12} \otimes X_{21}),
\end{eqnarray}
\end{widetext}
with $X_{ab}\equiv |a\ra \la b|$, $a = 1, 2$, being the Hubbard operators, and $|1 \rangle =  |\uparrow \rangle$, $|2 \rangle =  |\downarrow \rangle$, and $q = e^{ih}$ being the quantum deformation parameter, which can be  identified by comparing the eigenvalues of $\sigma = \hat{p} R$, obtained  to be $-q^{-3/2}$ and $\sqrt{q}$, with degeneracy $1$ and $3$, respectively, with the corresponding eigenvalues for the two-spin system in the correlated region, which can be computed, yielding $-e^{3\pi g/2}$ and $e^{-\pi g/2}$, respectively, resulting in $q = e^{-\pi g}$.

An $R$ matrix that satisfies the YBZ equation creates a quantum group/algebra~\cite{D-85,D-86,CGO-92}, by identifying the needed commutation relations, which in the case under considerations happens to be ${\rm SL}_{q}(2, \mathbb{C})$. Upon introducing a natural involution (complex conjugation) operation, it is reduced to its real ``compact'' version ${\rm SU}_{q}(2)$, which provides a representation theory with good properties, in particular, (i) decomposition of representations in irreducibles, (ii) adding two spins using ($q$-deformed) 3j-symbols, (iii) comparing the result of adding three spins in different order that gives rise to ($q$-deformed) 6j-symbols, (iv) there is a notion of unitary representations of ${\rm SU}_{q}(2)$, which makes the $q$-deformed 3-j and 6-j symbols scalar product preserving, and, very importantly, (v) the actions of the quantum group and the BG commute.


Summarizing, the quantum group ${\rm SU}_{q}(2)$ appears in the problem in a natural, yet not completely direct way. Indeed, the BCS problem possesses a spin interchange symmetry, which is preserved upon extension to the multi-time Schr\"odinger equation, and further identifies the braid group ${\rm B}_{N}$, rather than symmetric/permutation group ${\rm S}_{N}$ to describe the symmetry of the multi-time problem, which happens due to the poles of the $H_{j}$ Hamiltonians. It is the observation that the action of the BG on the states at $t\rar\infty$ naturally leads to an evolution matrix that satisfies the YBZ equation that defines a quantum group, and most importantly, whose action commutes with the BG, that brings in representation theory of quantum groups as a tool to analyze representations of ${\rm B}_{N}$ in the BCS problem, exactly in the same way as spin considerations allow the representation of the group of permutations to be analyzed, as described, e.g., in~\cite{LL-QM-3}.

\subsection*{Numerical solution of nonstationary Schr\"odinger equation}

\begin{figure} [!htb]
{\includegraphics[width=0.96\columnwidth]{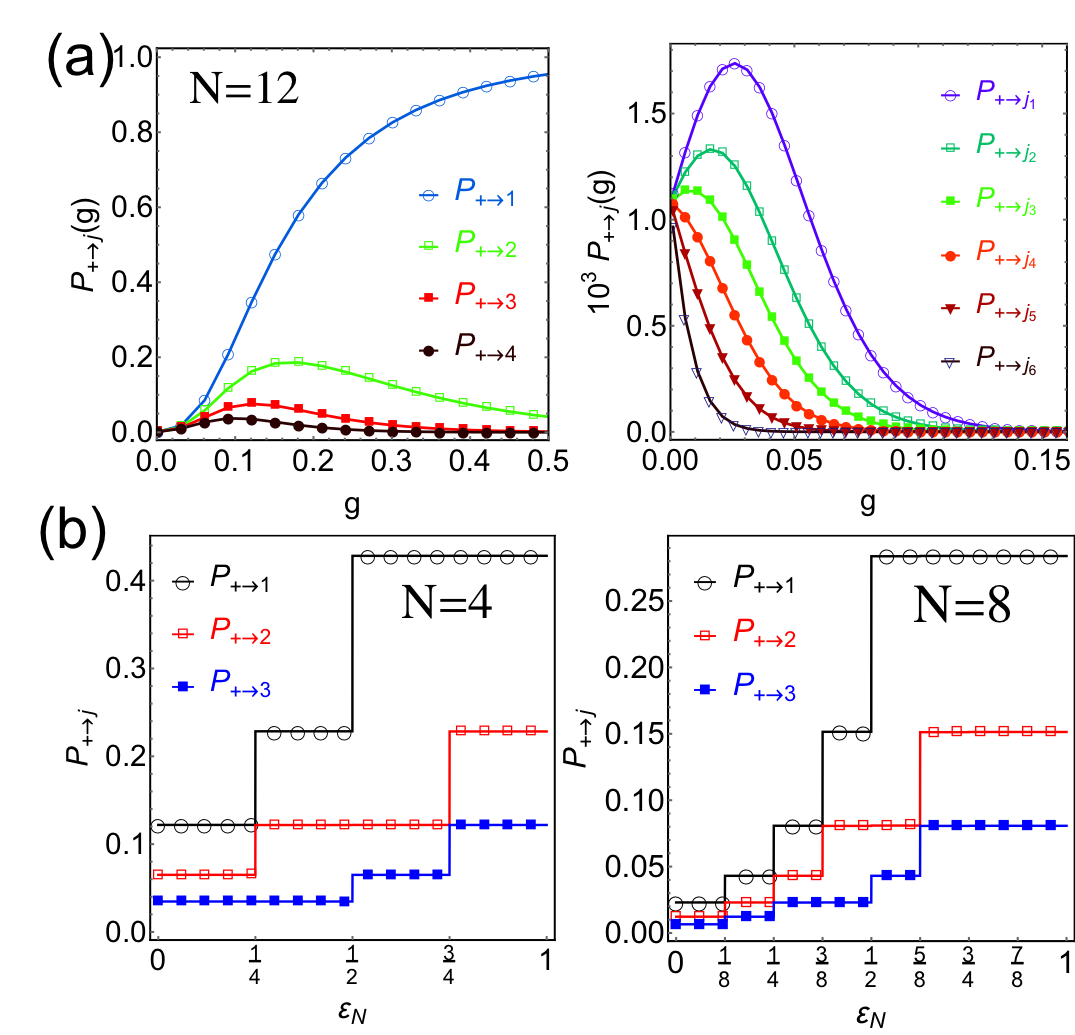}}
\caption{Transition probabilities at $S^{z}_{\rm tot}=0$ as functions of  {\bf (a)} coupling $g$ at $N=12$, and {\bf (b)} energy parameter $\ve_{N}$  at $g=0.1$. Evolution starts at $t=0.001$ in the ground state $|+\ra \equiv |\psi_0\ra$ (main text)  and ends at $t=1000$.  We always choose $\ve_j=j/N + \xi_j$, with $\xi_j$ being uniformly distributed random real numbers in the range $(-1/(2N), 1/(2N))$.  The microstates in (a)  are $|1\ra = |\ua\ua \ua \ua \ua \ua \da \da \da \da \da \da\ra$, $|2\ra = |\ua\ua \ua \ua \ua \da \ua \da \da \da \da \da\ra $,  $|3\ra = |\ua\ua \ua \ua \ua \da \da \ua  \da \da \da \da\ra $, $|4\ra = |\ua\ua \ua \ua \ua \da \da \da \ua \da \da \da\ra $, $|j_1 \ra = |\ua \ua \ua \da \da \da \ua \da \ua \da \ua \da\ra$, $| j_2 \ra = | \ua \da \ua \ua \da \da \ua \ua \da\da \da \ua  \ra$, $|j_3 \ra = | \ua\ua \ua\da \da \da \da \ua \da \da \ua \ua \ra$, $|j_4 \ra = | \ua \ua\da \da\da \da \ua \ua \da \ua\da \ua \ra$, $| j_5 \ra = |  \da \da\ua \da  \ua\ua \da \ua \da \da \ua \ua \ra $ and $|j_6 \ra = | \da\da\da\da\da\da\ua\ua\ua\ua\ua\ua \ra$. 
In (b), the final microstates for the left figure are $|1 \ra = | \ua \ua \da \da\ra$, $|2 \ra = | \ua  \da  \ua\da \ra$, and $|3 \ra = | \ua\da\da \ua  \ra$, and for the right figure: $| 1\ra = | \ua \ua \ua \ua \da \da \da \da\ra $, $|2\ra = | \ua \ua \ua \da \ua \da \da \da \ra$, and $|3\ra = | \ua \ua \ua \da \da \ua \da \da \ra$. 
Probabilities of final states depend only on the  relative order, but not on specific values of spin energy levels $\{ \ve_j \}$, as predicted by the exact solution. } \label{Fig:check2}
\end{figure}

Numerical simulations were performed by  solving  Schr\"{o}dinger equation 
\be
i \frac{\partial \Psi(t)}{\partial t} = \hat{H}_{\rm BCS}(t) \Psi(t),
\ee 
where $\hat{H}_{\rm BCS}(t)$ is the time dependent BCS Hamiltonian (main text) using  DSolve routine of Mathematica software.
 For the case of  $S^{z}_{\rm tot}=0$, we could simulate systems with up to $N=12$ spins, for which the Hilbert space dimension is $ C^N_{N/2}\sim 10^3$.  
  Figure~\ref{Fig:check2} shows perfect agreement  between numerical and theoretical, i.e. based on the  distribution in Eq.~(12) of the main text, predictions for transition probabilities to specific microstates.

\subsection*{Numerical comparison to a non-integrable model}
\begin{figure} [!htb]
{\includegraphics[width=1\columnwidth]{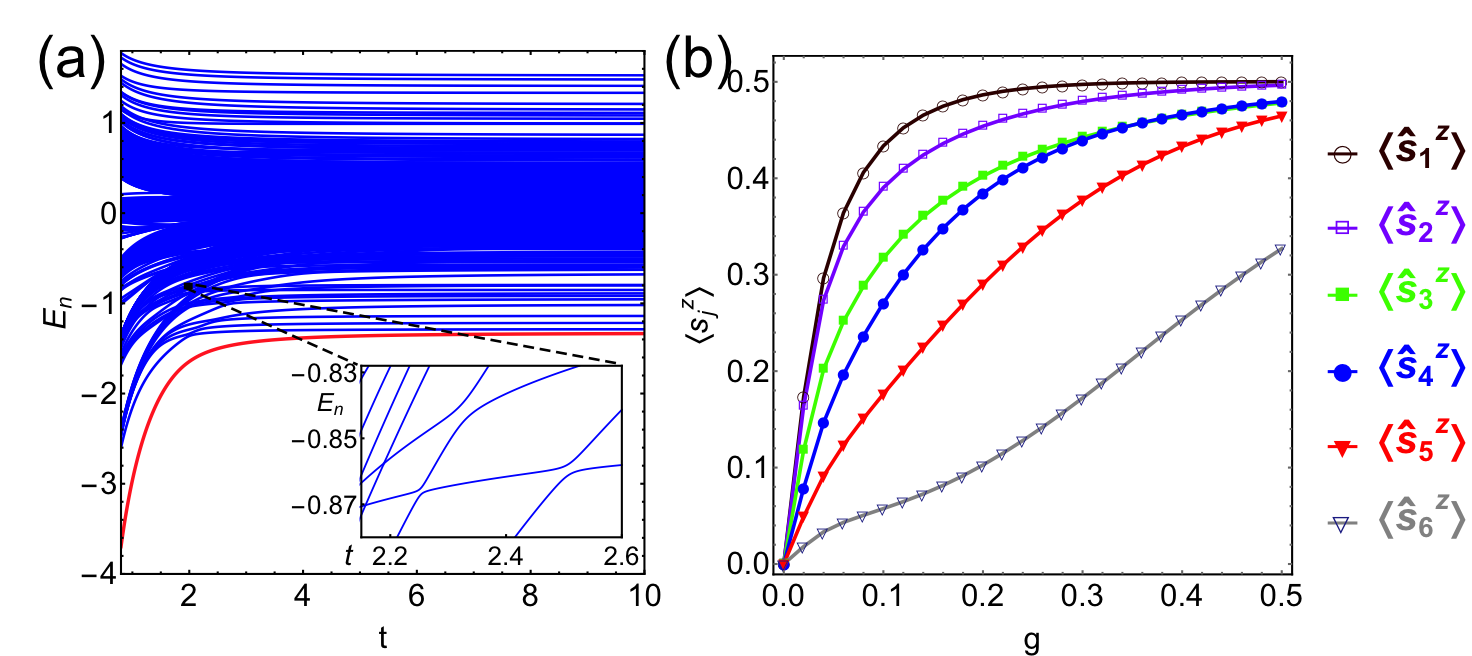}}
\caption{ {\bf (a)} The energy spectrum of the nonintegrable Hamiltonian (\ref{eq:H2}) with $g=1/N$ and $N=12$. The inset shows zoom-in energy levels with avoided crossings. {\bf (b)} Numerical results for the final polarization of several spins as functions of $g$ at $N=12$. Solid lines are guidelines for eyes. Couplings $J_{ijk}$ are the same for (a) and (b). They are chosen as $J_{ijk} = J_{i} J_j J_k$, with $J_{i}$ being a Gaussian distributed random number with zero mean and variance $\delta^2 J= 0.1$. } \label{Fig:triple}
\end{figure}

Numerical studies of nonintegrable QA models are extremely difficult due to many-body interactions and explicit time-dependence of the Hamiltonians. So, large-$N$ comparison with our BCS model is currently impossible. However, 
it is still instructive to look at models with numerically accessible numbers of spins. 

We considered
 the case with the same $\hat{H}_B$ and the same driving protocol but a more complex  Ising part of the Hamiltonian:
\be
\hat{H}' (t)= \sum_{i \neq j \neq k} J_{ijk} \hat{s}^z_i \hat{s}^z_j \hat{s}^z_k  - \frac{g}{t} \sum_{i \neq j} \hat{s}^{+}_i \hat{s}^{-}_j,  \label{eq:H2}\\ \nn
~~ i, j, k = 1, \ldots, N,
\ee
 where $J_{ijk}$ are randomly chosen couplings of simultaneously three spins.  This choice keeps basic original symmetries of the BCS model intact, so we can compare the previously used characteristics. 
Thus, the adiabatic energy spectrum in Fig.~\ref{Fig:triple}(a) 
shows that the energy level crossings are avoided, indicating that this model is no longer integrable. 

We solved the time dependent problem with the Hamiltonian (\ref{eq:H2}) numerically. Figure~\ref{Fig:triple}(b) shows the final average polarization of several spins for $N = 12$ and $S^z_{\rm tot} = 0$. 
Dependence of the polarization of individual spins on $g$ turns out to be qualitatively similar to the one in the BCS model (cf. Fig.~4(a) of the main text).
In particular, most of the spins find their ground state directions at the adiabaticity parameter values $g<1$, while only a couple  of spins in this case required $g>1$ regime in order to find their ground state. 

On the other hand, numerical results for the model (\ref{eq:H2}) showed also that simple estimates using the Landau-Zener formula, which we used to identify the onset of the adiabatic regime in the main text,  fail generally for some spins, so even at $g=1$ their polarization can be substantially different from the saturation value (as for the spin with index 6 in Fig.~\ref{Fig:triple}(b)). 
We attribute this behavior to the fact that the lowest energy excitations have more complex structure than simple two-spin flips in the BCS model. 

Based on our observations, we can speculate that, for larger $N$ values, it would be much harder to reach the adiabatic regime and
 find the exact ground state during QA in nonintegrable systems but calculations with a small error tolerance should be achievable considerably faster than during the precise QA.




\end{document}